\documentclass[10pt,letterpaper]{article}

\usepackage{cogsci}

\cogscifinalcopy 

\usepackage[T1]{fontenc}
\usepackage[american]{babel}
\usepackage{csquotes}
\usepackage{newtxtext,newtxmath}
\usepackage{graphicx}
\usepackage{dsfont}   
\usepackage{xcolor}
\usepackage{tikz}
\definecolor{ForestGreen}{HTML}{005f50}

\usepackage{amsmath}
\usepackage{booktabs}

\newtheorem{definition}{Definition}

\usepackage[
  backend=biber,
  style=apa,
  natbib=true,
  eprint=true,
  annotation=false,
]{biblatex}
\newcommand{\tA}{\texttt{A}}
\newcommand{\tE}{\texttt{E}}
\newcommand{\tI}{\texttt{I}}
\newcommand{\tO}{\texttt{O}}
\newcommand{\tac}{\texttt{ac}}
\newcommand{\tca}{\texttt{ca}}
\newcommand{\cE}{\mathcal{E}}
\newcommand{\R}{\mathbb{R}}
\newcommand{\EE}{\mathbb{E}}
\newcommand{\PP}{\mathbb{P}}

\newcommand{\Eins}{\mathbf{1}}

\usepackage{float} 

\usepackage[pdftitle={Response Variability and Stability in Human Reasoning}, citecolor=ForestGreen, urlcolor= ForestGreen, linkcolor= ForestGreen, colorlinks=true]{hyperref}	

\title{Response Variability and Stability in Human Reasoning}

\author[1]{\mbox{Clemens Bombach (clemens.bombach@hsw.tu-chemnitz.de)}}
\author[2]{\mbox{Rajmadan Lakshmanan}}
\author[3]{\mbox{Marco Ragni}}
\affil[1,2,3]{Predictive Analytics, TU Chemnitz}

\begin{document}

\maketitle

\begin{abstract}
Understanding how humans reason -- and how reasoning responses vary across tasks and individuals -- remains a core challenge for modeling and explanation in cognitive science. We investigate the stability of response patterns within reasoners and whether variation in these patterns can be used to predict learning effects. We introduce a formal, geometry-based method to quantify distances between individual reasoning patterns and their internal variability, grounded in heuristic theories. The proposed framework is tested against experimental data via generalized linear mixed-effects models and clustering, where we find that our proposed variation measure interacts with correctness to predict performance gains. Moreover, we find that reasoning patterns are stable over time within the same reasoner. The method is general enough to be applied to other reasoning domains.

\end{abstract}

\section{Introduction}
Many (classical) cognitive models implicitly assume an almost deterministic mapping from stimulus to response: holding the task and context constant, repeated presentations of the same input are expected to yield the same or at least similar response. This assumption simplifies model identification and parameter estimation, because variability can be treated as negligible. However, substantial trial-to-trial and inter-individual variability has been shown in reasoning and decision making, arising from factors such as lack of focus, attention, and context or sequence effects, often conceptualized as unwanted noise in decision making (e.g., \cite{kahneman2021noise}).

Consider the domain of human reasoning, more precisely the domain of syllogistic reasoning. A syllogism consists of two quantified premises over three terms, for example
\begin{center}
	\begin{tabular}{l}
		All Architects are Bakers, \\
		All Chemists are Bakers,
	\end{tabular}
\end{center}

and a conclusion such as ``All Architects are Chemists'' or ``Some Chemists are not Architects''. Normative and observed human responses in syllogistic reasoning differ substantially. For example, in a meta-analysis using data from 6 studies of all $64$ categorical syllogisms, it was reported that $47\%$ of all participants responded ``All Architects are Chemists'' \citep{Khemlani2012}. However, the logically correct response is that no valid conclusion can be derived from the premises, chosen by only $31\%$ of participants. None of the theories evaluated in the meta-analysis of \citet{Khemlani2012} adequately explain all gaps between normative and empirically observed responses, so the development of a general psychological theory remains an open problem. Interestingly, most of the 9 theories  reported assume a static cognitive process, i.e., for the same input of the task they predict in most cases the same response. 

Previous studies indicate a shift of the responses towards more logically correct responses if human reasoners are tested a week apart \citep{Ragni2018,Dames2022}. Responses in reasoning tasks do not vary at random, but instead reveal systematic patterns. Individual patterns are consistent across tasks \citep{stanovich1998individual}. Consistency judgments have been shown to be systematically influenced by underlying representations and processing constraints \citep{ragni2014evaluation}. 
However, with the exception of one theory (mReasoner), that change has not been explained by other theories and it has rarely been investigated on the individual level.
A related open question is to what degree small variations in the task change the response. For instance, the syllogism above is obtained from the syllogism
\begin{center}
	\begin{tabular}{l}
		All Architects are Bakers, \\
		All Bakers are Chemists,
	\end{tabular}
\end{center}
by swapping the terms in the second premise. Superficially, this is a small change, but it changes the normative response from ``All Architects are Chemists'' to ``No valid conclusion''. However, the results of \citet{Khemlani2012} indicates that most reasoners are insensitive to this change.

A major barrier in this regard is that there is no generally accepted formalism for determining whether a response pattern is more or less variable, or for defining the distance between two response patterns. Hence, this paper is concerned with developing such a formally sound framework. The principal idea is to identify individual reasoning patterns with \emph{functions} acting on spaces of binary vectors, thereby transforming questions about reasoning patterns into questions about functions on discrete spaces. Such functions are well studied in the mathematical literature, allowing us to draw on suitable notions of distance and variability.
This method of identification also incorporates findings from heuristic theories of syllogistic reasoning --- theories that predict responses from surface features of the premises alone, 
without evaluating logical validity --- including the atmosphere theory \citep{Woodworth1935}, Probability Heuristics Model~(PHM) \citep{chater_1999}, and TransSet \citep{Brand2019}.
More precisely, the atmosphere theory predicts conclusions from the universal/particular and positive/negative character of the premises.
The PHM ranks conclusions by informativeness derived from a probabilistic reading of the quantifiers, predicting that reasoner avoid uninformative conclusions (max-heuristic). 
TransSet models the reasoner's attempt to construct a transitive inference path across the figure of the syllogism.
These three theories together motivate our encoding of both tasks and responses into binary feature vectors, as detailed in the upcoming discussion.

We show the value of this framework by re-analyzing a dataset collected in a study on the stability of syllogistic reasoning over time \citep{Dames2022}. "Participants solved all 64 syllogisms twice, one week apart. 
Performance often improved in the retest, however the degree of improvement was highly dependent on the particular task and reasoner.
Based on this result and the discussion above, we formulate the following research questions:
\begin{itemize}
\item \textbf{Q0} What is a useful distance measure on syllogistic reasoning patterns? How can we measure response variability?
\item \textbf{Q1} To what extent does response variability explain  individual differences in performance improvement? In other words, are highly variable response patterns a predictor of retest performance?
\item \textbf{Q2} Related to the result of \citet{stanovich1998individual}: Are reasoning patterns stable within individuals? That is, is the distance between a reasoner's initial response pattern and their retest pattern smaller than the distance between their initial pattern and another reasoner's retest pattern?
\end{itemize}

\section{Materials and Methods}
We begin by briefly describing the dataset used for our analysis \citep{Dames2022}. This dataset was chosen because of its large sample size and because it includes repeated measures for all $64$ syllogisms. Session~0 assessed personality traits and general cognitive abilities. In Session~1, participants selected one response (out of the $9$ canonical responses) for each of the $64$ syllogistic problems; no feedback was provided, and task order was randomized. This procedure was repeated in Session~2, which took place one week after Session~1. The final sample consists of $N = 100$ individuals.

For the data analysis, we employ generalized linear mixed models (GLMMs) which were fit in R using the \texttt{lme4} package \citep{Bates2015}. GLMMs a are well-known tool for hierarchical modeling. For model diagnostics, we employed the \texttt{DHARMa} package \citep{Hartig2024}. For the calculation of $p$-values in Q2, we employed the \texttt{lmerTest} package \citep{Kuznetsova2017}. All models were fit using the BOBYQA optimizer. Goodness of fit was assessed with the \texttt{performance} package using pseudo-$R^2$ \citep{Luedecke2021, Nakagawa2013}.  The data and the scripts used to run the analyses are openly available on GitHub.\footnote{\url{https://github.com/c-bombach/cogsci-response-variability}}

The new ingredient of our approach is the systematic use of response variability and distance features, which allow a finer characterization of individual responses than correctness alone and are defined in the next section.

\section{Q0: Framework for Feature Sensitivity}

Feature sensitivity is operationalized here as local response variability with respect to a psychologically motivated task distance. A syllogism, as above, consists of two quantified premises about three terms $a$, $b$, $c$. Quantifiers are $\tA=\text{All}$, $\tE=\text{No}$, $\tI=\text{Some}$ and $\tO = \text{Some \dots not}$.

There are 4 possible orderings of the terms and these are called figures:
\begin{center}
	\begin{tabular}{cccc}
		Figure 1 & Figure 2 & Figure 3 & Figure 4 \\ 
		$a$-$b$ & $b$-$a$ & $a$-$b$ & $b$-$a$ \\
		$b$-$c$ & $c$-$b$ & $c$-$b$ & $b$-$c$ \\
	\end{tabular}
\end{center}

A syllogistic problem is uniquely determined by the quantifiers of the two premises and the figure (e.g., the syllogism from the introduction is $\tA \tA 3$), which we refer to as its \emph{surface features}. Analogously, responses can be encoded in terms of surface features, with \texttt{NVC} (No Valid Conclusion) as the logically correct 
response in $37$ of the $64$ syllogisms. The remaining $27$ syllogisms each license a valid conclusion, which must be one of $8$ possible response types consisting of a quantifier from $\{\tA,\tE,\tI,\tO\}$ combined with a direction $\{\tac, \tca\}$ (e.g., $\tE \tac$ for ``No $a$ are $c$'').

In the following, we denote by $X = \{(q_1,q_2,r): q_1, q_2 \in \{\tA,\tE,\tI,\tO\}, r \in \{1,2,3,4\} \}$ the set of syllogistic tasks. The set of responses is defined as $Y = \{(q,d): q \in \{\tA,\tE,\tI,\tO\}, d \in \{\tac,\tca\}\} \cup \{\texttt{NVC}\}$. As outlined in the introduction, we identify reasoning patterns with functions $f:X \to Y$ that take a task $x \in X$ as an input and produce an output $f(x) \in Y$.

To obtain notions of distance and variability for maps from $X$ to $Y$, we begin by defining distance functions (also called \emph{metrics}) between elements of the underlying spaces $X$ and $Y$ respectively. Formally, a metric on a set $Z$ is a function $d_Z:Z \times Z \to [0,\infty)$ such that  $d_Zf(z,w) = 0$ if and only if $z = w$, and for all $z,w,u \in Z$ it holds that  $d_Z(z,w) = d_Z(w,z)$ and $d_Z(z,u) \leq d_Z(z,w) + d_Z(w,u)$. The pair $(Z,d_Z)$ is called a metric space.

\subsection{Metrics on task and response sets}
A simple metric can be obtained by one-hot encoding. This means that each syllogism is encoded as a unit vector in $\R^{64}$. Then, we could take the Euclidean distance on $\R^{64}$. This, however, would make all syllogisms equidistant from each other. Moreover, in applications such as clustering, such an embedding suffers from the curse of dimensionality. By contrast, our goal is to obtain a metric to express response similarities and which lives on a low-dimensional space. If we wish to stick with a binary encoding, we need at least $6$ bits since $2^6 = 64$.

To motivate our definition of a metric on $X$, we recall that the syllogistic quantifiers can be divided into \emph{universal} quantifiers  $\tA,\tE$ versus \emph{particular} quantifiers $\tI, \tO$ and \emph{positive}  quantifiers $\tA,\tI$ versus \emph{negative} quantifiers $\tE,\tO$.
This classification is the basis of the atmosphere theory of syllogistic reasoning \citep{Woodworth1935}, which postulates that reasoners select conclusions according to the \emph{atmosphere} created by the premises.
Specifically, the theory predicts that the universality and positivity of the premises jointly determine the predicted conclusion: a single particular premise ($\tI$ or $\tO$) is sufficient to bias reasoners toward a particular conclusion, and a single negative premise ($\tE$ or $\tO$) toward a negative one.
Because conclusion selection is thus driven by these two binary features of each premise quantifier, encoding each quantifier along precisely the universality and positivity dimensions embeds the theoretically relevant structure directly into our metric.

We identify each quantifier $q$ with a tuple $(u_q,p_q)$ of bits $u_q,p_q$ that encode the universality and the positivity of $q$, respectively. More precisely, we have $u_q = 1$ if $q$ is universal and $0$ if it is particular. Analogously, we set $p_q = 1$ if $q$ is positive and $0$ otherwise.

In a similar fashion, we encode the figure $r$ using two bits $t_r,s_r$. In particular, we set $t_r = 1$ when $r \in \{1,2\}$. The definition of $t_r$ is motivated by the following considerations: In figure $1$ syllogisms, the ordering of terms allow for a simple transitive inference from $a$ to $c$ (even when this inference may not be logically warranted). A figure $2$ syllogism can be reduced to a figure $1$ syllogism by swapping the ordering of the premises. For the figures $3,4$, it is not immediately possible to obtain such a transitive path. We therefore see figure $1$ and $2$ syllogisms as \emph{transitive}. This distinction between transitive and intransitive figures plays an important role in TransSet \citep{Brand2019, Brand2020}, which is an algorithmic model of syllogistic reasoning. It is based on the  empirically supported observation that reasoners attempt to create a transitive path between inferences.
If such a path is not immediately available, they resort to a (possibly illicit) reordering of the terms.

If we wish to obtain a $6$-dimensional embedding, we are now left with one free bit $s_r$, that we set equal to $1$ if $a$ appears in the first premise. Thus, we set $s_r = 1$ if $r \in \{1,3\}$. We can therefore encode a syllogism $x = (q_1,q_2,r)$  as the $6$-dimensional binary vector
\begin{equation*}
	j(x) = (u_{q_1},p_{q_1},u_{q_2},p_{q_2},t_r,s_r) \in \{0,1\}^6 \,.
\end{equation*}
We obtain a one-to-one correspondence $j:X \to H_6$ between $X$ and the $6-$dimensional Hamming cube $H_6 = \{0,1\}^6$. On $H_6$, we have a natural metric given by the Hamming distance
\[
	d_{H_6}(v,w) = \sum_{m=1}^6 |v_m - w_m|\,.
\]
This counts the number of bits on which $v$ and $w$ disagree. We pull back this distance to a metric on $X$ using $j$ by setting $d_X(x,y) = d_{H_6}(j(x),j(y)).$

To obtain a metric on $Y$, we proceed in a similar fashion: Given $y \in Y$ we set $k(y) = (0,0,0,0)$ if $y = \texttt{NVC}$. Otherwise, if $y = (q,d)$ where $q \in \{\tA,\tE,\tI,\tO\}$ and $d \in \{\tac,\tca\}$, we set $k(y) = (u_q,p_q,1,0)$ if $d = \tac$ and $k(y) = (u_q,p_q,0,1)$ if $d = \tca$. Finally, we set
\[
	d_Y(y,\tilde y) = \sum_{j=1}^4 |k(y)_j - k(\tilde y)_j|\,.
\]
Geometrically, we can think of $k$ as mapping $Y$ injectively into a proper subset of the four-dimensional Hamming cube $H_4$.

Note that the \texttt{NVC}-response is mapped to the origin. Our choice of $k$ implies that $\tO$-responses lie closest to \texttt{NVC}.
This choice is inspired by the $\tO$-heuristic of the PHM theory of syllogistic reasoning \citep{chater_1999}.
PHM postulates that reasoners use heuristics that are based on the informativeness of the premises.
This informativeness is itself derived from a probabilistic interpretation of the quantifiers: $\tA$ is maximally informative as it rules out the most possibilities, 
while $\tO$ is least informative.
The $\tO$-heuristic states that reasoners generally avoid drawing $\tO$-conclusions due to their low informativity, suggesting a psychological similarity between \texttt{NVC} and $\tO$-conclusions.
By contrast, the most informative conclusions $\tA ac$ and $\tA ca$ have high distance to \texttt{NVC} in our embedding.
\subsection{Distance and variability of response patterns}
 
Having defined metrics $d_X,d_Y$ on $X$ and $Y$, we may now view reasoning patterns as maps $f:(X,d_X) \to (Y,d_Y)$ between metric spaces. On the set of these maps we consider, for $p \geq 1$, the $\ell^p(X;Y)$-metrics,
\begin{equation*}
    d_{\ell^p(X;Y)}(f,g) = \bigg( \sum_{x \in X} d_Y(f(x),g(x))^p \bigg)^{1/p}\,.
\end{equation*}
This is well-defined since $X$ is finite and generalizes the familiar Euclidean distance for $p=2$.

Since the Hamming cubes carry a natural graph structure, we may also see reasoning patterns as maps between graphs. This means we view $X,Y$ as graphs where the vertices $x,y$ are elements of $X,Y$ and two vertices are connected with an edge (written as $x \sim y$) if and only if the distance between them is $1$. Thus, the degree of a vertex in $X$ is given by $\deg(x) = |\{y: y \sim x\}| = |\{y: d(y,x) = 1\}|$.

We use this graph structure to define the desired similarity measure $f:X \to Y$. It is a generalization of the graph energy for real-valued functions on graphs (see, e.g., \cite{Keller2021}) to the setting where the target space $Y$ is a metric space:
\begin{definition}\label{def:energy}
	Let $p \geq 1$, $(X,d_X), (Y,d_Y)$ be metric spaces such that $X$ is the vertex set of a finite, connected graph. We define the local $p$-energy of $f:X \to Y$ at $x \in X$ as
	\[
	\cE_p(f;x) = \frac{1}{\deg(x)}\sum_{y \sim x}  d_Y(f(x),f(y))^p\,.
	\]
	The total energy of $f$ is defined as
	\[
	\cE_p(f) = \sum_{x \in X } \cE_p(f;x)\,.
	\]
    The respective normalizations are defined as
    \[
	E_p(f;x) = \cE_p(f;x)^{1/p}\,, \quad E_p(f) = \cE_p(f)^{1/p}\,. 
    \]
\end{definition}
Note that the local energy of $f$ at $x$ is $0$ if $f$ is constant on a neighborhood of $x$. Thus, the energy measures the degree to which $f$ is not (locally) constant and can therefore be seen as a measure of variability. Our definition is itself a special case of the notion of $p$-energy of maps between metric measure spaces as defined in the work by \citet{Jost1994}. Such energies have been used extensively in machine learning for the purpose of dimensionality reduction, for example in the Laplacian eigenmaps algorithm \citep{Belkin2003}.

\section{Meta-data inference}

In preparation for the predictive modeling, we perform a meta-data inference using $k$-means clustering. The dataset consists of repeated task--response trials collected from a fixed set of participants.
Let $I = \{1,\dots,M\}$ denote the set of participants, where each participant $i \in I$ completes multiple trials indexed by $t = 1,\dots,T$.
Each trial involves a syllogistic task $x_{i,t} \in X$, a response $f_i(x_{i,t}) \in Y$, and an associated local energy value derived from the response pattern.

The present analysis focuses exclusively on participant-level information.
This aggregation allows us to infer general response tendencies and stability in reasoning behavior.

For each trial $t$ of participant $i$, we define a feature vector $\mathbf{v}_{i,t} = (\mathbf{r}_{i,t},E_2(f_i; x_{i,t}))$
where
$\mathbf{r}_{i,t} \in \{0,1\}^{4}$ denotes the binary response-feature encoding.
$E_2(f_i; x_{i,t})$ denotes the normalized local $2$-energy introduced in Definition~\ref{def:energy}, evaluated at task $x_{i,t}$ for participant $i$.

To obtain a participant-level representation, we compute the empirical mean $\bar{\mathbf{v}}_i
=
\frac{1}{T}
\sum_{t=1}^{T}
\mathbf{v}_{i,t}$.
This vector summarizes the average response tendencies and energetic profile of participant $i$.
Before clustering, we standardize each feature dimension across participants to zero mean and unit variance.
This step ensures that task features, response features, and energy contribute comparably to the distance computations.

\begin{figure}[t]
    \centering
    \includegraphics[width=0.9\linewidth]{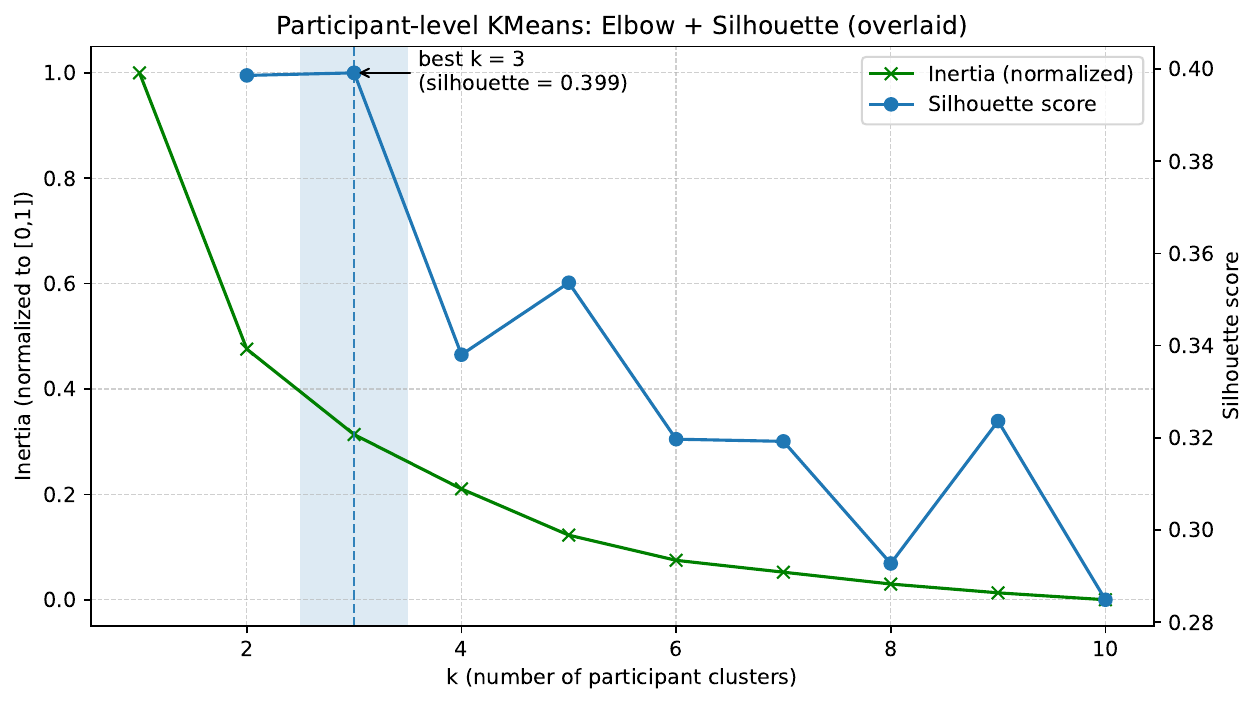}
    \caption{
Participant-level parameter ($k$) selection using the elbow and silhouette criteria.
The shaded region marks the selected number of clusters $k$.
}

    \label{fig:participant_kmeans_model_selection}
\end{figure}

We perform participant-level clustering using the $k$-means algorithm.
Let $\tilde{\mathbf{v}}_i$ denote the standardized participant-level feature vector.
For a fixed number of clusters $k$, the algorithm identifies cluster centroids
$\boldsymbol{\mu}_1,\dots,\boldsymbol{\mu}_k \in \mathbb{R}^n$
by minimizing the objective function
\[
\min_{\{c_i\},\,\{\boldsymbol{\mu}_j\}}
\sum_{i=1}^{N}
\left\|
\tilde{\mathbf{v}}_i - \boldsymbol{\mu}_{c_i}
\right\|^2,
\]
where $c_i \in \{1,\dots,k\}$ denotes the cluster assignment of participant $i$. Each centroid $\boldsymbol{\mu}_j$ represents a prototypical participant profile.
This profile captures a characteristic combination of task-feature exposure, response-feature usage, and energy measure.
To find the optimal number of clusters, we employ the elbow criterion and silhouette score. 
The elbow criterion measures how much the inertia, which measures within-cluster variability, decreases as the number of clusters increases.
On the other hand, the silhouette score quantifies how well-separated and coherent the resulting clusters are.
From Fig.~\ref{fig:participant_kmeans_model_selection}, both criteria indicate a clear balance at $k=3$, which we therefore select as the optimal number of participant clusters.

\begin{figure}[t]
    \centering
    \includegraphics[width=0.9\linewidth]{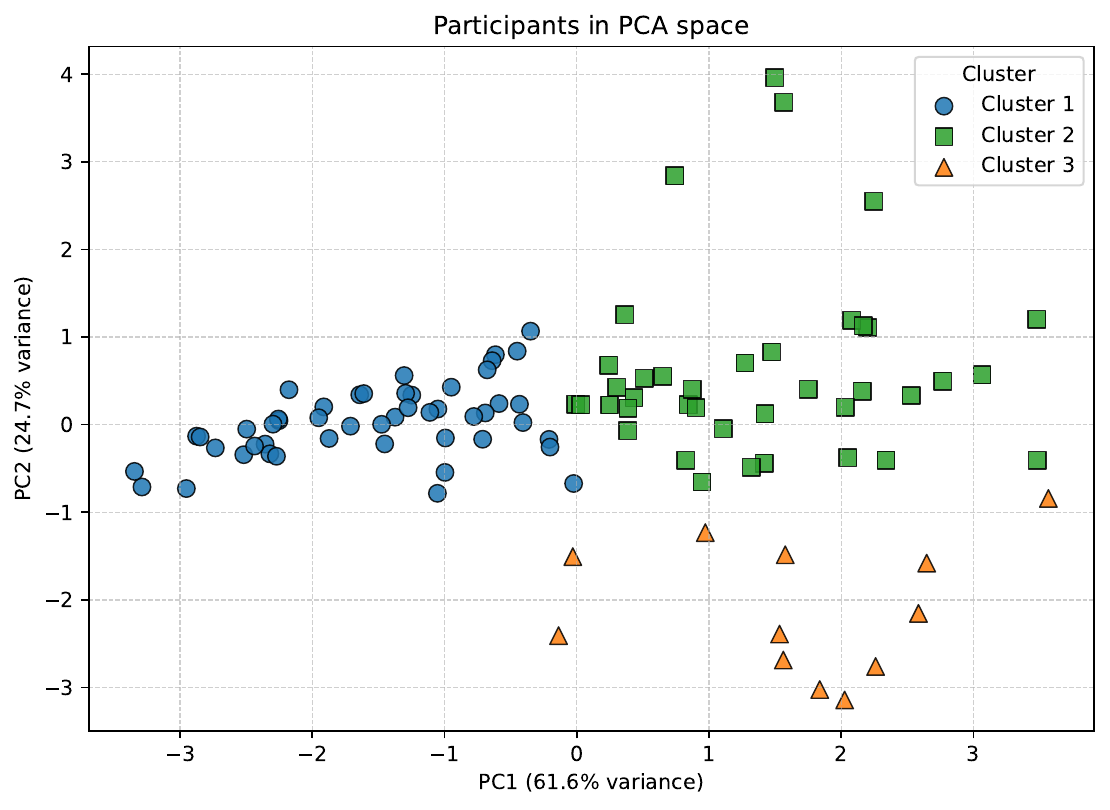}
    \caption{
Principal component analysis of participant-level feature vectors. Points represent participants projected onto the first two principal components; colors and marker shapes indicate $k$-means cluster membership. Axis labels show the variance explained by each component.
    }
    \label{fig:participant_pca}
\end{figure}

Participants in Cluster 1 were found to have substantially higher mean correctness ($67.2\%$) than participants in Clusters 2 ($37.9\%$) and 3 ($30.3\%$). By contrast mean energy was highest for Cluster 2 ($0.471$), followed by Cluster 3 ($0.389$) and Cluster 1 ($0.307$). We apply Principal Component Analysis~(PCA) after clustering to obtain a low-dimensional projection of participant-level feature vectors. Each participant is represented by averaged task features, response features, and local energy, so principal components reflect joint variation in strategy choice and response smoothness. Fig.~\ref{fig:participant_pca} shows the participant-level clustering projected onto the first two principal components. 

The PCA projection reveals clear structure across the three participant clusters. Principal components are influenced by both response tendency and energy, with the PC1 showing a stronger positive influence of energy, and PC2 being more influenced by response tendency, in particular with respect to direction (\texttt{ac} vs. \texttt{ca}). Participants in Cluster~1 occupy low values along the first principal component and form a relatively compact group along the second, indicating a \emph{coherent, correct response strategy} with low flexibility. Participants in Cluster~2 show high values along the first principal component and a broader spread along the second. Participants in Cluster~3 align with Cluster~2 along the first principal component but separate strongly along the second, occupying markedly lower values. This indicates differences in the preferred response direction between the two clusters.

\section{Mixed-effects modeling}
\subsection{Q1: Prediction of retest performance}
For this research question, we studied the effect of the energy in the initial test on correctness in the retest. This can be reformulated as follows: When reasoners give a response that is exceptional in the sense that it has a large distance from responses to neighboring syllogisms (in the sense of the metric defined above), does this indicate a performance improvement when they are tested again? We fit a sequence of GLMMs with logit link function with the aim of predicting retest correctness $C_{\text{retest}}$ from the following variables:
\begin{itemize}
    \item initial correctness $C$, encoded such that $C=1$ for correct responses and $C= 0$ for incorrect responses
    \item task validity $V$, encoded such that $V=1$ for valid tasks and $V = -1$ for invalid tasks
    \item participant-centered and scaled, normalized local $2$-energy, $E_2$
    \item (standardized) correctness per task averaged over all participants $\overline C$
\end{itemize}
In particular, we were interested in the effect of the energy variable and its interactions.

Continuous variables were standardized to aid convergence and to make results easily interpretable. We decided to use the normalized $2$-energy since it corresponds to the Dirichlet energy. The normalization reduces skewness and ensures that it is in the same ``length unit'' as the underlying spaces. We opted for participant-centering since we are interested in predicting the effect of local deviations in energy on learning. Mean correctness $\overline C$ was included as a predictor since models that didn't include it showed substantial underdispersion during model diagnostics. We presume that this is caused by the large variation in task success rate since some syllogisms (such as $\tA\tA1$) are solved correctly by almost all participants.

To find a suitable model, our aim was to find a convergent model with a maximally complex random effects structure and minimal AIC. The best model we were able to fit under these criteria is given by the following formula in Wilkinson-Rogers notation \citep{Wilkinson1973}:
\begin{align}\label{eq:full_model}
    C_{\text{retest}} \sim &C*V*E_2 - C:V:E_2 + \overline C \notag \\
    &+ (C +V |\, \text{participant}) + (C + E_2 |\, \text{task}) \tag{$M_1$}
\end{align}
In order to have a baseline for assessing the effect of energy, we fitted the reduced model
\begin{equation}\label{eq:red_model}
    C_{\text{retest}} \sim C*V +\overline C 
    + (C +V |\, \text{participant}) + (C|\, \text{task}) \tag{$M_0$}\,.
\end{equation}

A model comparison showed a significant improvement in fit of \eqref{eq:full_model} over \eqref{eq:red_model} ($\Delta \mathrm{AIC} = 39, G^2(6) = 50.82, p = 3.2\cdot10^{-9}$). This strongly indicates that energy has substantial predictive value for the problem under consideration. Overall, model fit was acceptable with $R^2_{\text{conditial}} = 0.644, R^2_{\text{marginal}} = 0.382$. %
The results for the fixed effects are listed in Table \ref{tab:best_model}. As expected, correctness and mean correctness have the largest effects on retest correctness with coefficients $\beta_C = 1.560$ and $\beta_{\overline C} = 1.307$ respectively. The effects of validity and energy are smaller ($\beta_V = 0.242, \beta_{E_2} = 0.225$) but the low standard error for energy in particular $(\sigma_{E_2} = 0.076)$ suggests a robust finding. That improvement in correctness is higher for valid syllogisms is consistent with the findings of \citet{Dames2022}. We note the significant correctness-energy interaction $\beta_{C:E_2} = -0.438$.

\begin{table}[t]
\caption{Fixed effects for models $M_1$ and $M_0$. $N=6400$ observations; 100 participants; 64 tasks.}
\centering
\begin{tabular}{lcc}
\toprule
 & $M_1$ & $M_0$ \\
\midrule
Intercept                & $-0.467^{**}\,(0.152)$  & $-0.360^{*}\,(0.141)$ \\
Correctness $C$          & $1.560^{***}\,(0.139)$  & $1.446^{***}\,(0.133)$ \\
Validity $V$             & $0.242^{*}\,(0.121)$    & $0.227^{*}\,(0.109)$ \\
Energy $E_2$       & $0.225^{**}\,(0.076)$   &  \\
Av. Corr. $\overline C$  & $1.307^{***}\,(0.074)$  & $1.281^{***}\,(0.073)$ \\
$C:V$                    & $0.037\,(0.121)$        & $-0.018\,(0.111)$ \\
$C:E_2$                  & $-0.438^{***}\,(0.100)$ &  \\
$V:E_2$                  & $-0.122\,(0.064)$       &  \\
\midrule
AIC                      & $5442.222$              & $5481.051$ \\
BIC                      & $5577.503$              & $5575.748$ \\
Log Likelihood           & $-2701.111$             & $-2726.526$ \\
\bottomrule
\multicolumn{3}{l}{\scriptsize{$^{***}p<0.001$; $^{**}p<0.01$; $^{*}p<0.05$}}
\end{tabular}
\label{tab:best_model}
\end{table}
To get a better sense of the effect of energy on performance improvement, it is interesting to compute effective odds ratios dependent on the value of the correctness and validity variables for energy in \eqref{eq:full_model}:
		\begin{center}
			\begin{tabular}{lrr}
            \toprule
				$OR(E_2)$ & valid & invalid\\
				\midrule
				correct & 0.72 & 0.91 \\
				incorrect & 1.11 & 1.41 \\\bottomrule
			\end{tabular}
		\end{center}
In particular, we have $OR < 1$ for correct and $OR > 1$ for incorrect responses. 
One possible interpretation for these effects is as follows: When a correct response has high energy, it is different from responses to related syllogisms and thus it is less likely that it can be associated with a consistent strategy of the participant. On the other hand, low energy can be taken as a sign of a systematic response pattern. Thus, syllogisms with correct but unsystematic responses are less likely to be answered correctly on retest. By contrast, for incorrect responses, high energy raises the probability of improvement. In particular, this effect is stronger for incorrect responses to invalid syllogisms.

As a sanity check, we replaced the theory-driven encoding based on task features by a simple one-hot encoding. As expected, this led to a substantial reduction in fit over the best model with theory-driven encoding ($\Delta \mathrm{AIC} = - 25)$ and made the main effect of energy non-significant ($\beta = 0.070$). However, we note the observation of a significant validity-energy interaction ($\beta = -0.25, p < 0.0001$). A possible explanation is that a preference towards \texttt{NVC}-responses leads to a large subset on which the reasoning pattern is constant. This lowers the energy in an encoding-independent fashion. Such an effect can therefore be detected by a coarse encoding.

\subsection{Q2: Stability of reasoning patterns}
\begin{figure}[t]
  \begin{center}
    \includegraphics[width =0.4\textwidth]{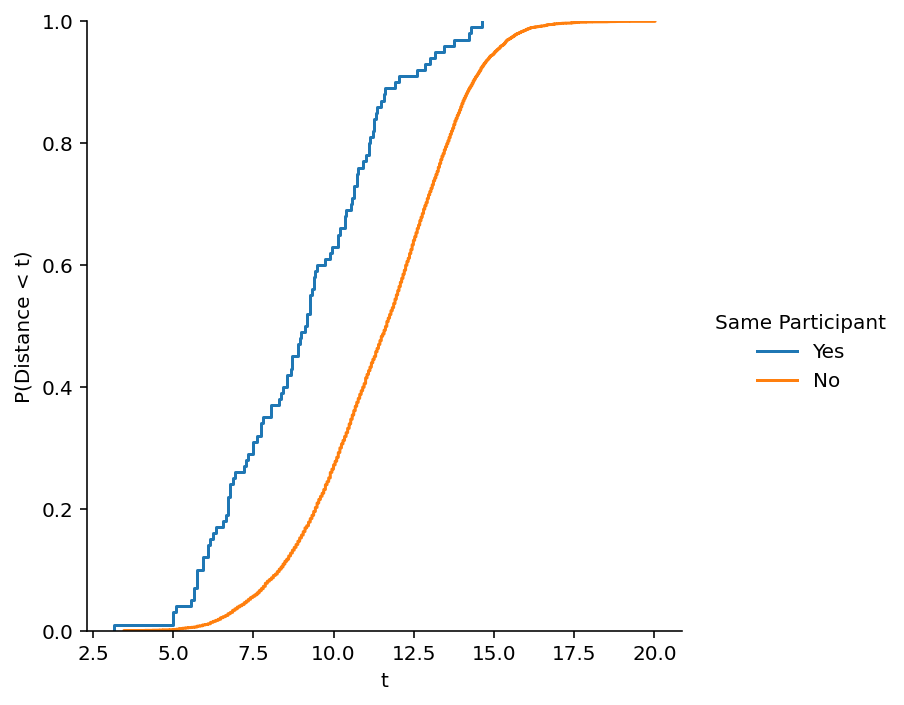}
  \end{center}
  \caption{Empirical cumulative distribution functions (cdfs) for test-retest distance.}
  \label{fig:cdf}
\end{figure}

Formally, the question is: Suppose that $i,j$ are reasoners, $f^i_0,f^j_0$ their response patterns on the initial test and $f^i_1,f^j_1$ their response patterns on the retest. Moreover, let us write $d_{ij} = d_{\ell^2(X;Y)}(f^i_0,f^j_1)$ in the following. If response patterns are stable within reasoners, then we should have
\begin{equation}\label{eq:cond_exp}
    \EE(d_{ij}| i = j) < \EE(d_{ij}| i \neq j)
\end{equation}
with $\EE$ denoting the (conditional) expectation. In other words, the expected distance of a reasoner between a test and retest pattern is smaller when both patterns come from the same reasoner.
A plot of the empirical conditional distribution functions suggests that this is indeed the case. In Figure~\ref{fig:cdf}, the blue curve (same participant) is shifted to the left of the orange curve (different participants), indicating that within-reasoner distances are stochastically smaller than between-reasoner distances. Formally, our sample fulfills the stronger condition $\PP(d_{ij} < t| i = j) > \PP(d_{ij} < t| i \neq j)$ for all $t > 0$.
In order to estimate the effect size, we fit a LMM of the form
\begin{equation*}
    d_{ij} \sim \Eins_{i = j} + (1|i) + (1|j)\,,
\end{equation*}
where $\Eins_{i = j}$ denotes a binary variable that is $1$ if and only if $i = j$ and $0$ otherwise. The terms $(1|i), (1|j)$ are random intercepts for the participants $i,j$. We obtained a (fixed) intercept of $\alpha = 11.48$ and a fixed effect of $\beta = -2.44$ distance units with a standard deviation $\sigma =0.13$, resulting in the 95\% confidence interval $[-2.69, -2.19]$, indicating a very significant and strong effect. We note that $R^2_{\text{conditional}} = 0.709$ whereas $R^2_{\text{marginal}} = 0.011$, suggesting that while the effect itself is strong, it only explains a small part of the total variation in distance. This is to be expected since reasoning patterns differ greatly between individuals.

\section{Discussion}
We developed a method to define psychological distances between individual reasoning patterns, incorporating findings from heuristic theories of reasoning. We employed a notion from metric geometry, the so-called $p$-energy (Definition \ref{def:energy}), as a measure of response variability. The framework supports fine-grained analyses of response variability at the level of individual reasoner--task combinations. The clustering analysis suggests that reasoners differ systematically both in response strategies and variability.

Our findings from the mixed-effects modeling are twofold: First, we showed that high energy predicts an increase in correctness on retest, but only for initially incorrect responses. For initially correct responses, higher energy predicts a lower probability of a correct retest response. Energy can thus be seen as a measure of psychological consistency in response patterns: If reasoners are stably incorrect, an improvement is less likely than when they are unstably incorrect. Second, we demonstrated that individual differences in reasoning patterns are stable: reasoners were closer to their own retest patterns than to other reasoner's retest patterns. This effect seems to be particularly robust.

We note that our framework abstracts away from content effects such as belief bias \citep{Khemlani2012, chater_1999}.
However, the task formulations in the dataset under consideration here are specifically designed to minimize content effects. Thus, this limits the impact on the present results.
More broadly, extra-logical factors such as content effects and belief bias represent additional sources of response variability that the present framework does not capture; extending the task encoding to incorporate them would be a natural refinement.

Regarding generalizability, some aspects of our construction depend on the domain of syllogistic reasoning and the psychological theories adopted; alternative distance functions could be derived from mental model or logic-based approaches. However, the mathematical framework itself is domain- and theory-independent: defining distances on task and response spaces directly induces a distance and energy functional on response patterns. Recent Bayesian approaches to syllogistic reasoning predict probability distributions over responses \citep{tessler2022}. Extending our framework from deterministic response functions to non-deterministic response distributions would therefore be a natural next step and would accommodate experimental designs allowing multiple responses, which are natural for valid syllogisms licensing several conclusions.

\section{Acknowledgements} The authors thank the German Research Foundation (DFG) for financial support via grants awarded to MR within the projects 283135041 and 427257555. CB was funded by the Deutsche Forschungsgemeinschaft (DFG, German Research Foundation) -– Project number 283135041. RL was funded by the Deutsche Forschungsgemeinschaft (DFG, German Research Foundation) -– Project numbers 283135041; 427257555.

\printbibliography

\end{document}